\documentstyle[wstwocl,epsfig]{article}
\begin{document}

\bibliographystyle{unsrt}    % for BibTeX - sorted numerical labels

\newcommand{\st}{\scriptstyle}
\newcommand{\sst}{\scriptscriptstyle}
\newcommand{\mco}{\multicolumn}
\newcommand{\epp}{\epsilon^{\prime}}
\newcommand{\vep}{\varepsilon}
\newcommand{\ra}{\rightarrow}
\newcommand{\ppg}{\pi^+\pi^-\gamma}
\newcommand{\vp}{{\bf p}}
\newcommand{\ko}{K^0}
\newcommand{\kb}{\bar{K^0}}
\newcommand{\al}{\alpha}
\newcommand{\ab}{\bar{\alpha}}
\newcommand {\pom}  {I\hspace{-0.2em}P}
\def\be{\begin{equation}}
\def\ee{\end{equation}}
\def\bea{\begin{eqnarray}}
\def\eea{\end{eqnarray}}
\def\CPbar{\hbox{{\rm CP}\hskip-1.80em{/}}}%temp replacement due to no font

\def\ap#1#2#3   {{\em Ann. Phys. (NY)} {\bf#1} (#2) #3.}
\def\apj#1#2#3  {{\em Astrophys. J.} {\bf#1} (#2) #3.}
\def\apjl#1#2#3 {{\em Astrophys. J. Lett.} {\bf#1} (#2) #3.}
\def\app#1#2#3  {{\em Acta. Phys. Pol.} {\bf#1} (#2) #3.}
\def\ar#1#2#3   {{\em Ann. Rev. Nucl. Part. Sci.} {\bf#1} (#2) #3.}
\def\cpc#1#2#3  {{\em Computer Phys. Comm.} {\bf#1} (#2) #3.}
\def\err#1#2#3  {{\it Erratum} {\bf#1} (#2) #3.}
\def\ib#1#2#3   {{\it ibid.} {\bf#1} (#2) #3.}
\def\jmp#1#2#3  {{\em J. Math. Phys.} {\bf#1} (#2) #3.}
\def\ijmp#1#2#3 {{\em Int. J. Mod. Phys.} {\bf#1} (#2) #3}
\def\jetp#1#2#3 {{\em JETP Lett.} {\bf#1} (#2) #3.}
\def\jpg#1#2#3  {{\em J. Phys. G.} {\bf#1} (#2) #3.}
\def\mpl#1#2#3  {{\em Mod. Phys. Lett.} {\bf#1} (#2) #3.}
\def\nat#1#2#3  {{\em Nature (London)} {\bf#1} (#2) #3.}
\def\nc#1#2#3   {{\em Nuovo Cim.} {\bf#1} (#2) #3.}
\def\nim#1#2#3  {{\em Nucl. Instr. Meth.} {\bf#1} (#2) #3.}
\def\np#1#2#3   {{\em Nucl. Phys.} {\bf#1} (#2) #3}
\def\pcps#1#2#3 {{\em Proc. Cam. Phil. Soc.} {\bf#1} (#2) #3.}
\def\pl#1#2#3   {{\em Phys. Lett.} {\bf#1} (#2) #3}
\def\prep#1#2#3 {{\em Phys. Rep.} {\bf#1} (#2) #3.}
\def\prev#1#2#3 {{\em Phys. Rev.} {\bf#1} (#2) #3}
\def\prl#1#2#3  {{\em Phys. Rev. Lett.} {\bf#1} (#2) #3}
\def\prs#1#2#3  {{\em Proc. Roy. Soc.} {\bf#1} (#2) #3.}
\def\ptp#1#2#3  {{\em Prog. Th. Phys.} {\bf#1} (#2) #3.}
\def\ps#1#2#3   {{\em Physica Scripta} {\bf#1} (#2) #3.}
\def\rmp#1#2#3  {{\em Rev. Mod. Phys.} {\bf#1} (#2) #3}
\def\rpp#1#2#3  {{\em Rep. Prog. Phys.} {\bf#1} (#2) #3.}
\def\sjnp#1#2#3 {{\em Sov. J. Nucl. Phys.} {\bf#1} (#2) #3}
\def\spj#1#2#3  {{\em Sov. Phys. JEPT} {\bf#1} (#2) #3}
\def\spu#1#2#3  {{\em Sov. Phys.-Usp.} {\bf#1} (#2) #3.}
\def\zp#1#2#3   {{\em Zeit. Phys.} {\bf#1} (#2) #3}

\setcounter{secnumdepth}{2} % Number sections and subsections

%%%%%%%%%%%%%%%%%%%%%%%%%%%%%%%%%%%%%%%%%%%%%%%%%%
%                                                %
%    BEGINNING OF TEXT                           %
%                                                %
%%%%%%%%%%%%%%%%%%%%%%%%%%%%%%%%%%%%%%%%%%%%%%%%%%

\title{SOFT INTERACTIONS AT HIGH ENERGY}

\firstauthors{Aharon Levy}

\firstaddress{DESY, Hamburg, Germany\\and\\Tel--Aviv University, Tel--Aviv,
Israel}

\secondauthors{\  }
\secondaddress{\  }

\twocolumn[
\begin{flushleft}
\tt DESY 95-204 \\
November 1995\\
\end{flushleft}

\vspace{2cm}

\maketitle

\vspace*{12cm}

\thefootnote{\large{Rapporteur talk given at the International Europhysics
Conference on High Energy Physics, Brussels, July 27--Aug 2, 1995}}
]
\thispagestyle{empty}
\setcounter{page}{0}

\newpage

\twocolumn[
\vspace{18cm}
]
\thispagestyle{empty}
\setcounter{page}{0}

\newpage

\title{SOFT INTERACTIONS AT HIGH ENERGY}

\firstauthors{Aharon Levy}

\firstaddress{DESY, Hamburg, Germany\\and\\Tel--Aviv University, Tel--Aviv,
Israel}
\secondauthors{\  }
\secondaddress{\  }

\twocolumn[\maketitle
\vspace{-1.5cm}
\abstracts{
Soft interactions are not easily disentangled from hard ones. In an
operational definition of soft and hard processes one finds that at
presently analyzed scales there is an interplay of soft and hard
processes. As the scale increases, so does the amount of hard processes.
So far, nothing is as soft nor as hard as we would like.}]

\section{Introduction}

Soft interactions are usually understood as the interactions of hadrons at
a relatively small scale. It is sometimes also referred to as low $p_T$
physics, choosing the transverse momenta $p_T$ involved in the process as
the representative scale. Clearly this definition is very vague, as the
meaning of low $p_T$ is time dependent. We will try to give an operational
definition at least for some cases, how one could define more precisely
soft and hard interactions.

One could take the approach that anything that can be calculated by pQCD
can be called a hard process. All the rest would be soft. The problem
however is that what we calculate and what we measure is not the same. In
fact, we calculate processes between partons. These `dress--up' before
becoming the colourless object that we measure and thus introduce
elements like fragmentation and hadronization which we don't know to
calculate in QCD. We thus have an interplay of soft and hard processes,
which will be the main motto of this talk.

In the past, hadron--hadron interactions were thought off as the main
source of information about soft interactions, while $e^+e^-$
interactions and leptoproduction were traditionally designed to study
point--like interactions. The inverse seems to be happening today. The
hadron colliders are used for producing the point--like objects while the
others provide information on soft phenomena.

The talk will contain two main parts. The first part starts with a short
discussion about total cross sections since the bulk of the processes
making up the total cross section are believed to be soft interactions.
There were however some predictions that the total photoproduction cross
section in the HERA energy range will increase sharply due to the
resolved part of the photon structure, a prediction which was not borne
out by experiment. New results about the photon structure function will
be presented from the LEP experiments, which constrain the quarks in the
photon.

To learn about the gluon density distribution in the photon one studies
jet production at HERA. However, for understanding jets one
needs to learn about fragmentation and hadronization and jet shapes. The
LEP experiments have submitted very detailed studies about that and about
event shapes, particle rates and Bose--Einstein correlations, some of
which will be reviewed in this talk.

HERA studies have shown that when viewed from the Breit frame, the current
region in $ep$ interactions is very similar to that of $e^+e^-$ reactions,
so one can profit from all the studies made at LEP. However the proton
side at HERA is still not understood so well. There are indications that
much tuning has still to be done in the Monte Carlo generators for a good
description of the proton region.

The second part of this talk will be devoted to the interplay between
soft and hard interactions by giving some operational definition of soft
and hard for the most inclusive (total cross section) and most exclusive
(elastic cross section) processes. Are deep inelastic scattering (DIS)
processes soft or hard? This question will be discussed by studying the
behaviour of the total $\gamma^*p$ cross section with energy.

One of the surprising results of HERA was the existence of large rapidity
gap events in DIS. These showed the behaviour of a diffractive process in
which a pomeron is exchanged and which would traditionally be considered
as a soft phenomenon. Is that the case also in diffractive reactions in
DIS? In a picture where the pomeron is a pseudo particle which contains
partons, the Ingelman--Schlein pomeron, the large rapidity gap events in
DIS carry information about the structure of the pomeron. Further
information about the pomeron can be obtained from exclusive vector meson
production in $\gamma p$ and in $\gamma^* p$ reactions. Does one get a
consistent picture from these studies?

Finally, some preliminary evidence for the existence of colour--singlet
exchange at high $t$ will be presented.

\section{Cross sections}

\subsection{Total cross section}

Donnachie and Landshoff~\cite{DL} succeeded to describe all available
$\bar{p} p$, $p p$, $K^{\pm} p$, $\pi^{\pm} p$ and $\gamma p$ total cross
section values by a simple parametrization of the form $\sigma_{tot} = X
s^{0.0808} + Y s^{-0.4525}$, where $s$ in the square of the total center
of mass energy and $X$ and $Y$ are parameters depending on the interacting
particles. The value of $X$ is constrained to be the same for particle and
antiparticle beams to comply with the Pomeranchuk theorem~\cite{Pom}. The
power of the first term, which in the Regge picture is connected to the
intercept of the exchanged pomeron at $t$ = 0, ($\alpha_{\pom}(0)$ =
1.08), was determined by using the $\bar{p} p$ total cross section
measurement of E710~\cite{E710} at $\sqrt{s}$ = 1.8 TeV.  CDF~\cite{CDF}
repeated this measurement at the same $\sqrt{s}$ and found a significantly
higher value for the cross section, which would imply a higher pomeron
intercept of $\alpha_{\pom}(0)$ = 1.11.  Unfortunately, this discrepancy
between the two measurements has not yet been resolved.

Though an $ep$ collider is mainly viewed as a machine to study DIS, the
bulk of the neutral current (NC) cross section is in events where the
exchanged particle is a photon with a low virtuality $Q^2$. By using a
calorimeter in the electron direction which measures scattered electrons
at very small angles ($<$ 5 mrad), one can tag events produced by photons
of virtuality $Q^2 <$ 0.02 GeV$^2$, with a median $Q^2$ of $10^{-5}$
GeV$^2$. This way one can study photoproduction reactions of almost real
photons at center of mass energies of $\sim$ 200 GeV, which is an order
of magnitude higher than was previously available.

\setlength{\unitlength}{0.7mm}
\begin{figure}[hbtp]
\begin{picture}(100,110)(0,1)
\mbox{\epsfxsize8.0cm\epsffile{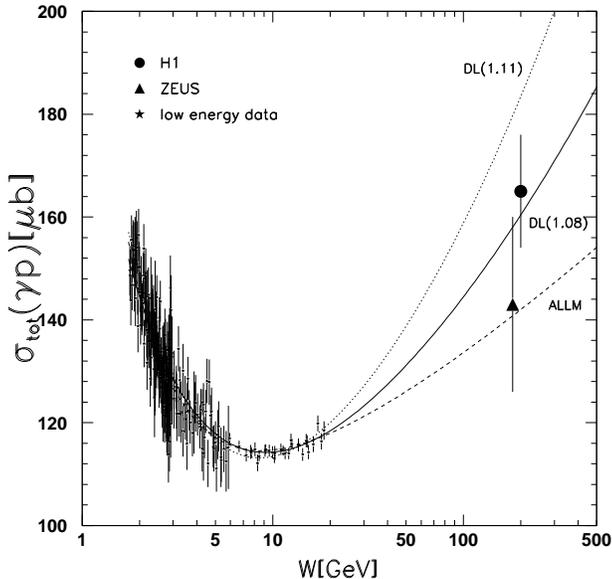}}
\end{picture}
\caption{Total photoproduction cross section as a function of the $\gamma p$
center of mass energy $W$ compared to some parametrization, as described
in the text. }
\label{fig:h1tot}
\end{figure}

The H1~\cite{H1stot} collaboration submitted a new measurement of the total
photoproduction cross section at a $\gamma p$ center of mass energy $W$ of
200 GeV. Their value of $\sigma_{tot}(\gamma p) = 165 \pm 2 \pm 11 \mu$b,
plotted in figure \ref{fig:h1tot} together with previous
measurements~\cite{Zstot},
agrees well with the value of 1.08 for the pomeron intercept and is lower
than that expected from an intercept of 1.11. In addition, this
measurement is higher than that expected from the ALLM~
\cite{ALLM} parametrization, to
be discussed later.

\subsection{Elastic and diffractive cross section}

The H1 collaboration measured also cross sections for elastic and diffractive
photoproduction processes. In photoproduction reactions, the elastic process
is defined through the reactions $\gamma p \to V p$, where $V=\rho^0$,
$\omega$, and $\phi$.  Assuming that the double dissociation cross section is
15 $\mu$b, the following results are obtained: $\sigma_{el}(\gamma p) \equiv
\sigma(\gamma p \to V p)$ = 17 $\pm$ 4 $\mu$b and $\sigma_{SD}(\gamma p)
\equiv \sigma(\gamma p \to X p) + \sigma(\gamma p \to V Y)$ = 32 $\pm$ 12
$\mu$b. The single diffraction (SD) processes contain those in which the
photon diffracts into a system $X$ and also the process in which the photon
turns into one of the three vector mesons $\rho^0$, $\omega$, or $\phi$ and
the proton diffracts into a system $Y$. One can then express these results as
ratios to the total cross section:

\begin{eqnarray}
\frac{\sigma_{el}}{\sigma_{tot}} & = & 0.10 \pm 0.03 \nonumber \\
\frac{\sigma_{SD}}{\sigma_{tot}} & = & 0.19 \pm 0.07
\label{ratio}
\end{eqnarray}
These results are in good agreement with those obtained by the
ZEUS~\cite{Zstot}
collaboration at $W$ = 180 GeV, which are 0.13 $\pm$ 0.05 for the elastic
ratio and 0.19 $\pm$ 0.04 for the SD ratio.

\setlength{\unitlength}{0.7mm}
\begin{figure}[hbt]
\begin{picture}(100,110)(0,1)
\mbox{\epsfxsize8.0cm\epsffile{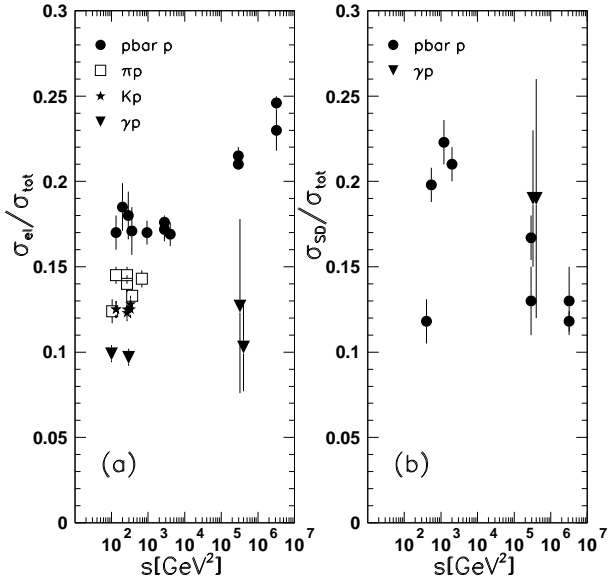}}
\end{picture}
\caption{A compilation of the ratio of elastic to total cross section for
$\bar{p} p$, $K p$, $\pi p$ and $\gamma p$ reactions as function of the
square of the center of mass energy $s$. }
\label{fig:ratio}
\end{figure}

It is interesting~\cite{GLM} to compare these results to a compilation of
$\bar{p} p$, $p p$, $K p$, $\pi p$ cross sections shown in figures
\ref{fig:ratio}(a) and \ref{fig:ratio}(b).  The ratio of the elastic to total
cross section of $K p$ and $\pi p$ reactions, wherever available, are smaller
than the ratio for $\bar{p} p$ ones, as expected. The $\gamma p$ ratio seems
to be smaller than one would naively expect, given the hadronic nature of the
photon. On the other hand, from the comparison in figure \ref{fig:ratio}(b),
the ratio of single diffractive to total processes seems to be higher than
expected.  One possible explanation of this fact can be connected to the way
one defines the elastic reaction in photoproduction as that of the three
lightest vector mesons. Though this might have been reasonable at low
energies, it may be underestimating the process at higher energies where the
available energy can produce many more vector mesons. Thus, at higher energy
one would move some of the cross section contained in the SD to the elastic
channel, increasing the elastic ratio and decreasing the single diffractive
one.

In spite of this open question of the exact definition of the subprocesses
in photoproduction, the total photoproduction cross section has no
ambiguity and its energy dependence is in accord with expectations from a
soft pomeron phenomenology. It doesn't show the dramatic rise which was
predicted by some so--called mini--jet models in which the contribution to
the rise comes from the partonic picture of the photon, discussed
in the next chapter.

\section{The photon structure function}

\subsection{$F_2^{\gamma}$ from $e^+e^-$ reactions}

The photon is the elementary gauge particle responsible for
electromagnetic interactions. Nevertheless, when interacting with hadrons,
it behaves as though it consists of two parts: one in which it acts as a
structureless elementary `bare' photon and a second part in which the
photon acts almost like a hadron~\cite{Greview}. Thus one can talk about
the notion of the photon structure function and measure it experimentally.
It is not exactly like a hadron because the `structured' part of the
photon also consists of two parts: one where the photon acts as a
hadronlike object and the process involves non--perturbative effects. This
part calls for a treatment similar to that of the proton and is thus
denoted as $F_2^{HAD}$. In the second part, which is unique to the photon,
the photon splits into a $q\bar{q}$ pair, and we will denote it by
$F_2^{box}$.  This part is also called the point--like part of the photon
or the anomalous part of the photon structure function and can be exactly
calculated through the process $\gamma \gamma \to \gamma \gamma$ described
by a box diagram.

For the hadronic part of the photon structure function one has
the same situation as for a regular hadron in the sense that one needs a
certain input at a starting scale $Q_0^2$. However the hope was that if
one can decompose the photon structure function in these two parts
\begin{equation}
F_2^{\gamma} = F_2^{box} + F_2^{HAD}
\end{equation}
then by going to high enough $Q^2$, $F_2^{\gamma} \to F_2^{box}$.
Since this part is exactly calculable and depends only on the QCD scale
parameter $\Lambda_{QCD}$, one could determine the QCD scale parameter by
comparing the calculation to the measured structure function.

It was however realized that one can not neglect the hadronic part even at
high $Q^2$ because this part is needed in order to cancel some
singularities at small Bjorken $x$. The situation is not improved by using
next to leading order (NLO) calculations. Thus one analyses the photon
structure function in a similar way to that of the proton structure
function by assuming an input parton distribution at a given scale and
evolving to higher scales. Another approach~\cite{FKP} is to use separate
inputs for the point--like and the hadronlike components of the photon
structure function with a cutoff parameter $p_T^0$ to separate the two
components.  A compilation~\cite{f2gcom} of $F_2^{\gamma}$ averaged in the
region $0.3 < x < 0.8$ can be seen in figure \ref{fig:f2gamma}.

\setlength{\unitlength}{0.7mm}
\begin{figure}[hbt]
\begin{picture}(100,110)(0,1)
\mbox{\epsfxsize8.0cm\epsffile{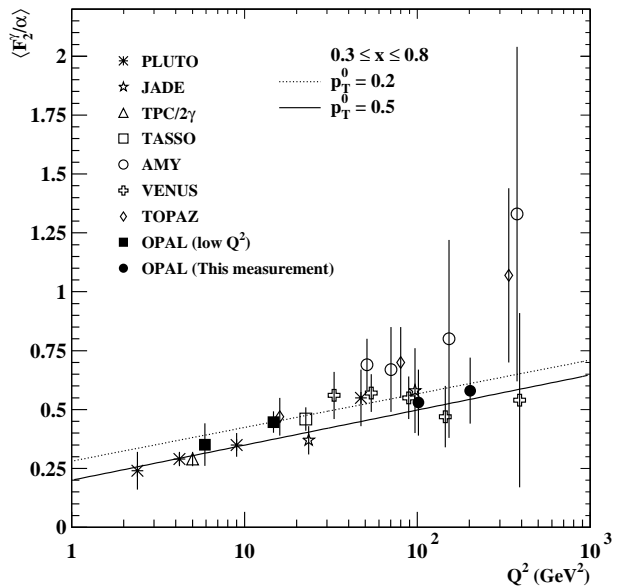}}
\end{picture}
\caption{A compilation of the photon structure function averaged in the
region $0.3 < x < 0.8$. The lines correspond to two values of the cutoff
parameter $p_T^0$.}
\label{fig:f2gamma}
\end{figure}

The measurement of the photon structure function is an experimental
challenge, especially in the low $x$ region. One has the problem of
unfolding the true $x$ distribution from the value of the visible $x$
measured in the detector. This unfolding seems to depend on the models
used in the Monte Carlo generators and further studies are needed. The
OPAL~\cite{f2gcom} collaboration is currently investigating this
difficulty.  The DELPHI~\cite{DELf2g} collaboration has attempted
a measurement of the photon structure function down to the
lowest $x$ value published so far, $x=0.001$. This result, if confirmed,
seems to indicate that the structure function does not rise at low $x$,
contrary to the behaviour of the proton structure function.

The different parametrizations of the parton distributions in the photon
are all obtained by fitting their expected evolution with $Q^2$ to the
results of the measured $F_2^{\gamma}$. These measurements constrain only the
quark distributions in the photon, which is why the gluon distributions
are so strongly dependent on which parametrization one uses. In order to
constrain also the gluons in the photon one needs to get information from
HERA measurements.

\subsection{Parton distributions in the photon from HERA}

The two components of the photon, the `bare' photon and the photon with
structure, are denoted in leading order (LO) by direct photon and
resolved photon, respectively, and are depicted in figure \ref{fig:direct}.

\setlength{\unitlength}{0.7mm}
\begin{figure}[hbt]
\begin{picture}(100,50)(0,1)
\mbox{\epsfxsize8.0cm\epsffile{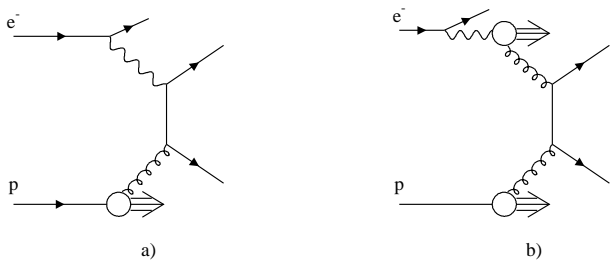}}
\end{picture}
\caption{Examples of leading order diagrams for (a) direct and (b)
resolved photoproduction.}
\label{fig:direct}
\end{figure}

In diagram (a) the whole momentum of the photon takes part in the
reaction with the parton of the proton resulting in two high $p_T$ jets.
In diagram (b), only part of the photon momentum participates in the
production of the two jets. Denoting by $x_{\gamma}$ the fraction of the
photon momentum participating in the hard process, one expects
$x_{\gamma} \sim 1$ for the direct photon and $x_{\gamma} < 1$ for the
resolved photon process. One way of estimating $x_{\gamma}$ is by
defining an observable variable $x_{\gamma}^{OBS}$ from the measured high
$p_T$ jets which would be a good approximation of the $x_{\gamma}$
calculated from the two final state partons:
\begin{equation}
x_{\gamma}^{OBS} = \frac{E_T^{j1}e^{-\eta^{j1}} + E_T^{j2}e^{-\eta^{j2}}
}{2E_{\gamma}}
\end{equation}
where $E_T$ is the transverse energy of the jet, $\eta$, its pseudo
rapidity, and $E_{\gamma}$ is the energy carried by the almost real
photon. The distribution~\cite{Zxg} of this variable is seen in figure
\ref{fig:xg}.
\setlength{\unitlength}{0.7mm}
\begin{figure}[hbt]
\begin{picture}(100,120)(0,1)
\mbox{\epsfxsize8.0cm\epsffile{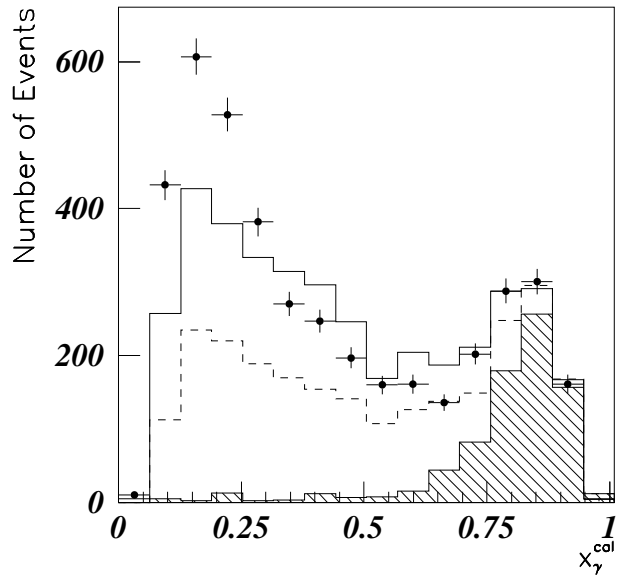}}
\end{picture}
\caption{The $x_{\gamma}$ distribution. The solid circles are uncorrected
ZEUS data. The solid (dashed) line represents the distribution
from the PHYTIA (HERWIG) simulation. The LO direct contribution to the
HERWIG distribution is shown by the shaded histogram. The Monte Carlo
curves have been normalized to fit the direct peak in the data.}
\label{fig:xg}
\end{figure}
The clear peak at high values of $x_{\gamma}^{OBS}$ indicates the presence
of direct type processes while the resolved photon processes populate the
lower $x_{\gamma}$ values.

\begin{figure}[hbt]
\begin{picture}(100,70)(0,1)
\mbox{\epsfxsize8.0cm\epsffile{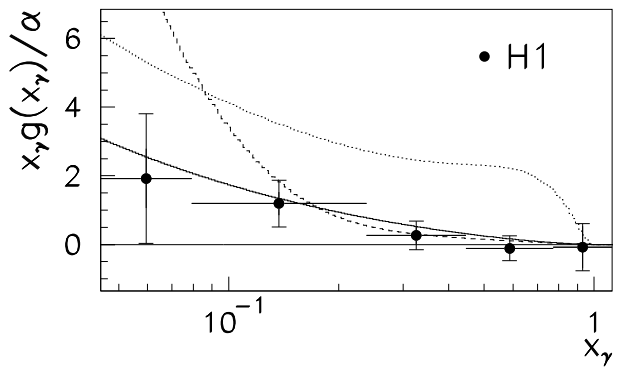}}
\end{picture}
\caption{ The gluon density distribution in the photon. The data is
represented by the solid dots. The lines are expectations from different
parametrizations of the gluon distribution in the photon.}
\label{fig:h1xg}
\end{figure}
The $x_{\gamma}$ distribution can in principle be used to extract information
about the gluon distribution in the photon. One can assume that the
distribution in figure \ref{fig:xg} is a sum of the direct and resolved
photon processes. By subtracting the estimated direct photon contribution one
is left with that coming from the resolved photon processes.  This part
consists of the contribution from quarks and from gluons. One can estimate
the quark part using a parametrization of the quark distribution function as
determined from fits to the photon structure function. Thus the difference
between the data and the quark estimate can be attributed to the gluons in
the photon~\cite{H1xg}. This difference can be converted into a gluon
distribution function of the photon, shown in figure \ref{fig:h1xg} .
The data are given at a scale of $ < p_T >^2 =$ 75 GeV$^2$. The lines are
expectations from different parametrizations of the gluon distribution in
the photon.

The method described above gives a rough estimate and a more careful
study is planned. However it shows the potential of the high $E_T$ jets
measured in the photoproduction data of HERA to provide information about
the gluons in the photon. The preliminary results given above do not
favor a steeply rising gluon distribution at low $x_{\gamma}$ predicted
by some of the parametrizations.

\subsection{Structure of virtual photons}

Measurements of the photon structure function has been done so far for
real photons. The measurement of $x_{\gamma}$ at HERA presented above has
been done for almost real photons having virtualities with a median of
$10^{-5}-10^{-3}$ GeV$^2$. Do photons of higher virtuality also have
structure? The only measurement so far of a structure function of virtual
photons has been done more than 10 years ago by the PLUTO~\cite{PLUTO}
collaboration for a virtuality of 0.35 GeV$^2$, showing a slight decrease
of the structure function compared to the one for real photons.

\begin{figure}[hbt]
\begin{picture}(100,110)(0,1)
\mbox{\epsfxsize8.0cm\epsffile{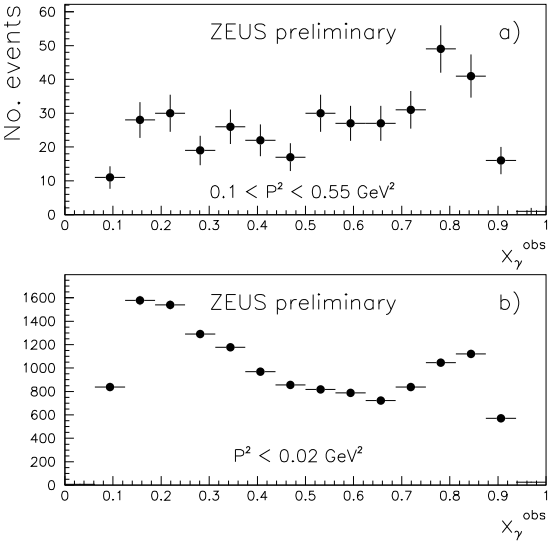}}
\end{picture}
\caption{Uncorrected $x_{\gamma}^{obs}$ distributions for a) virtual and
b) quasi-real photons.}
\label{fig:xgams}
\end{figure}
\begin{figure}[hbt]
\begin{picture}(100,110)(0,1)
\mbox{\epsfxsize8.0cm\epsffile{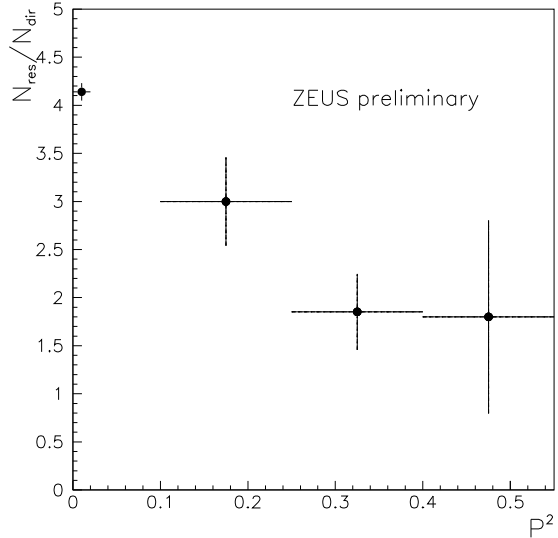}}
\end{picture}
\caption{The uncorrected ratio $N_{res}/N_{dir}$ as a function of photon
virtuality $P^2$ (in GeV$^2$). }
\label{fig:resdir}
\end{figure}

The ZEUS detector implemented a special beam--pipe calorimeter (BPC) which
can tag photons of virtuality $ 0.1 < P^2 < 0.55$ GeV$^2$. In $\gamma
\gamma$ physics the virtuality of the target photon is denoted by $P^2$
and that of the probing photon is denoted by $Q^2$. The uncorrected
$x_{\gamma}$ distribution is shown in figure \ref{fig:xgams} for the
measurements in the BPC~\cite{Zvirt} and compared to the distribution
measured in the luminosity electron tagger which measures almost real
photons.  One sees from figure \ref{fig:xgams}a that in addition to the
peak near $x_{\gamma}^{obs} \sim$ 1, which indicates the presence of direct
photon processes, there are events also at lower $x_{\gamma}^{obs}$. One
thus can conclude that photons of virtuality in the range $ 0.1 < P^2 <
0.55$ GeV$^2$ also have a resolved part. Furthermore, by making an
operational definition of direct photons as those with $x_{\gamma}^{obs}
>$ 0.75, one can study the ratio of the resolved to direct photon
processes as a function of the photon virtuality $P^2$, shown in figure
\ref{fig:resdir}.  This ratio should not be affected strongly by the
acceptance corrections.
The ratio of resolved to direct photon processes seems to decrease with
increasing photon virtuality in the $P^2$ range measured here.

The discussion in these last two subsections shows that in order to be
able to reconstruct $x_{\gamma}$ and thus gain information about parton
distributions in the photon at HERA one needs a good understanding of
jets. We therefore turn now to the next chapter where we describe the
present status of our knowledge about fragmentation and hadronization
studies.

\section{Fragmentation and hadronization studies}

Much of the material in this chapter comes from the LEP experiments and
have been covered in the parallel sessions by J.Fuster, W.Metzger,
R.Settles, T.Sj\"ostrand and F.Verbeure. Thus I bring here only a short
selection to highlight some points.

\subsection{Jet shapes}

\begin{figure}[hbtp]
\begin{picture}(100,150)(0,1)
\mbox{\epsfxsize8.0cm\epsffile{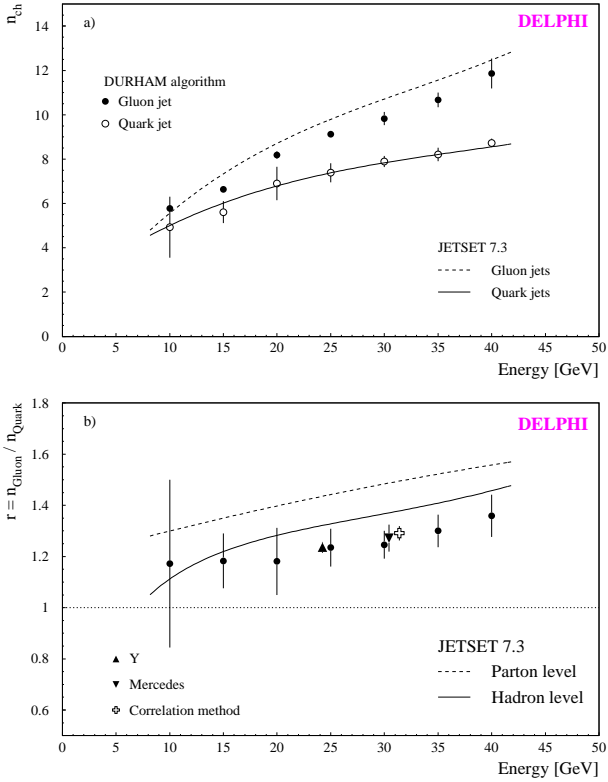}}
\end{picture}
\caption{(a) The gluon and quark charged multiplicity distribution as
function of the jet energy. (b) The ratio of gluon to quark multiplicities
as function of the jet energy.  }
\label{fig:gmult}
\end{figure}

As was shown in the earlier section, one needs to reconstruct jets in
order to get information about the momentum of the partons participating
in the hard processes. It is therefore important to have a good
understanding of the jet shapes, which are influenced by fragmentation
models. Properties of quark jets have been studied for quite some time,
thus there is a need to learn more about gluon jets. The
DELPHI~\cite{DELjet} collaboration presented a study of the energy
dependence of the difference between the quark and gluon jet
fragmentation. Using quark and gluon jets of purities $\sim$90\% they show
that gluon jets are broader than quark ones. They measured the charged
multiplicity in each as function of the jet energy (figure
\ref{fig:gmult}a) and also the dependence of the ratio between the gluon
and quark charged multiplicities on the energy (figure \ref{fig:gmult}b).
The results were compared with the expectations from the generator
JETSET7.3.  Not unexpectedly, the description of the gluon jet is not
perfect, firstly because it was much less studied and secondly maybe because
it has a more complicated colour structure than the quarks. For example,
because of QCD coherence, the gluon radiation angle has to be taken into
account properly~\cite{Dokshitzer}.

\subsection{Event shapes, particle rates}

{}From jet shapes, we turn to the description of multiparticle final states.
The ALEPH~\cite{ALEnch} collaboration studied the charged multiplicity
distribution in different rapidity intervals. They unfolded the data and
studied the charged multiplicity in a small rapidity range, a medium range
and over the whole rapidity window, as indicated in figure
\ref{fig:nobsaleph}. The JETSET7.3 predictions seem to be the only ones
which can give a good description of the data in all the rapidity
intervals.

\setlength{\unitlength}{0.7mm}
\begin{figure}[hbtp]
\begin{picture}(100,150)(0,1)
\mbox{\epsfxsize8.0cm\epsffile{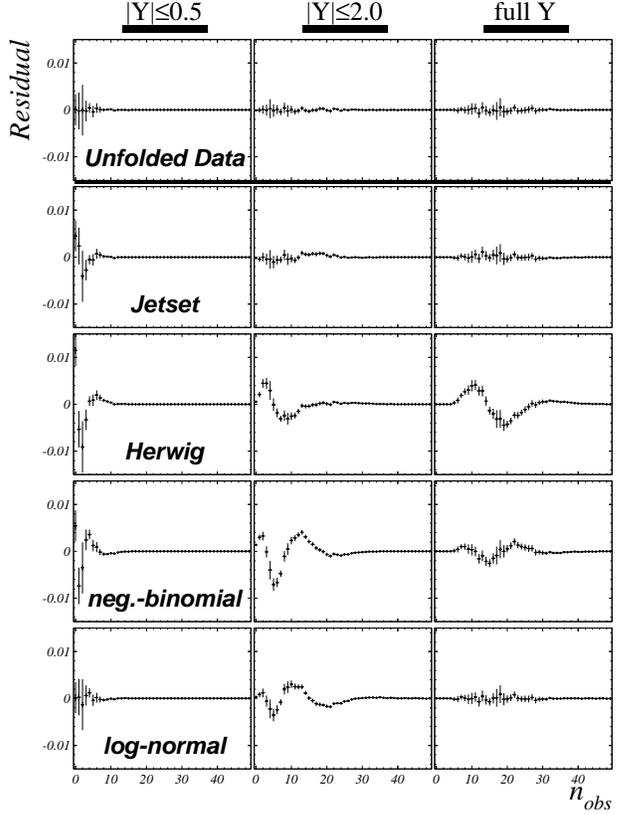}}
\end{picture}
\caption{Comparison of charged multiplicity data with unfolded results and
with
predictions of four models in small, medium and full rapidity range. }
\label{fig:nobsaleph}
\end{figure}

It is interesting to check whether the hadronic decomposition of the final
state is in agreement with those expected from the generators. One can get
the hadrons in the final state either as being formed directly from the
string or as decays of more massive hadrons. The OPAL~\cite{OPALhad}
collaboration studied the dependence of $\xi_{peak}$ on the hadron mass,
where $\xi_{peak}$ is the peak value of the variable $\xi=\ln(1/x_p)$ with
$x_p$ being the scaled hadron momentum. As one can see from figure
\ref{fig:hadron} when one includes in addition to the directly produced
hadrons also those from the decay of heavier hadrons, the JETSET7.3 Monte
Carlo reproduces the pattern observed in the data.

\setlength{\unitlength}{0.7mm}
\begin{figure}[hbtp]
\begin{picture}(100,100)(0,1)
\mbox{\epsfxsize7.0cm\epsffile{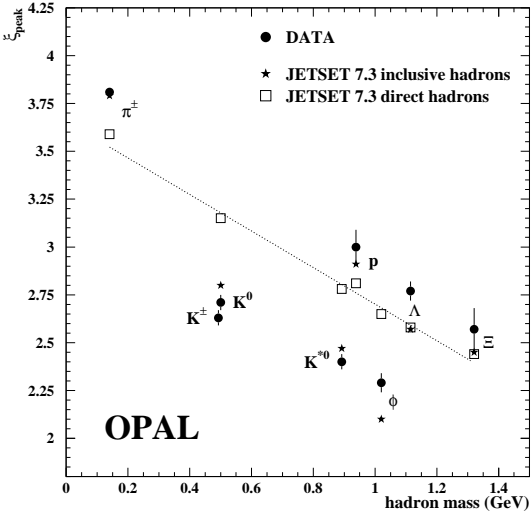}}
\end{picture}
\caption{The dependence of $\xi_{peak}$ on the hadron mass, compared to
prediction from the JETSET Monte Carlo for direct and inclusive production
of particles.}
\label{fig:hadron}
\end{figure}

In order to reproduce the rate of inclusive strange particle production,
the OPAL collaboration altered the default value of the strangeness
suppression factor $\lambda_s$ from 0.3 to 0.245. The ALEPH~\cite{ALEstr}
collaboration finds that it can describe the strange vector meson rate
using the default value of $\lambda_s$=0.3. At HERA, both the
ZEUS~\cite{ZEUSstr} and the H1~\cite{H1str} collaborations find that the
data are better described with a strangeness suppression value of
$\lambda_s \sim$ 0.2.

The DELPHI~\cite{DELtun} collaboration made a very detailed study of
tuning parameters of fragmentation models based on identified particles.
They succeed to find parameters which give a good description of most
features of the data. The one worrying result is that all models
underestimate the tail of the $p_T^{out}$ distribution by more than 25\%,
where $p_T^{out}$ is the transverse momentum of a particle out of the
event plane (see figure \ref{fig:ptout}). From the study of all other
variables, the authors conclude that the best overall description is
provided by ARIADNE4.06.

Such studies~\cite{DELtun} are most useful and should be encouraged since
they increase our ability to isolate which of the observations come from
fragmentation and which from the underlying basic process.

\setlength{\unitlength}{0.7mm}
\begin{figure}[hbtp]
\begin{picture}(100,120)(0,1)
\mbox{\epsfxsize7.0cm\epsffile{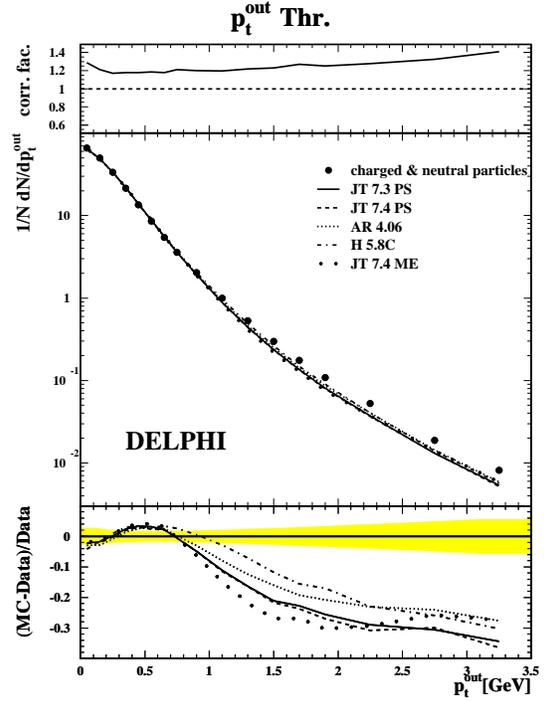}}
\end{picture}
\caption{Comparison of $p_T^{out}$ with respect to the Thrust axis with
different Monte Carlo models.}
\label{fig:ptout}
\end{figure}

Can one apply the results obtained at LEP to the HERA physics? There are some
clear similarities between $e^+e^-$ and $e p$ reactions. The best way to see
that is to compare the $e^+e^-$ results with those of $e p$ obtained in the
Breit frame.
\setlength{\unitlength}{0.7mm}
\begin{figure}[hbtp]
\begin{picture}(120,130)(0,1)
\mbox{\epsfxsize7.0cm\epsffile{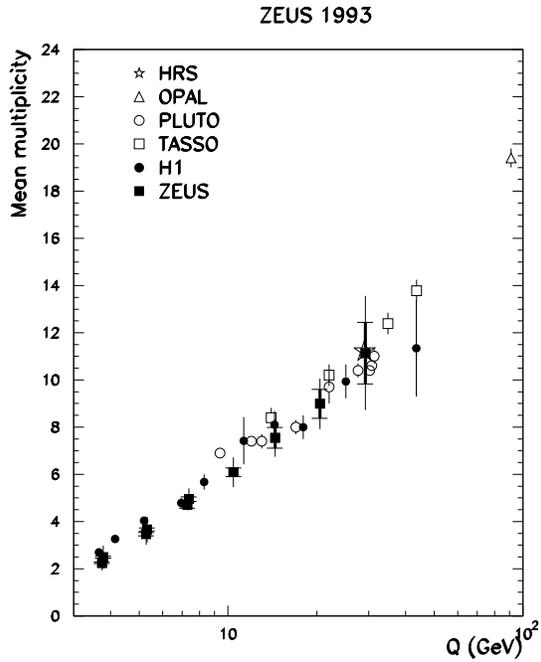}}
\end{picture}
\caption{Twice the mean charged multiplicity of $e p$ reactions in the Breit
frame compared to the mean charged multiplicity in $e^+e^-$ reactions.}
\label{fig:nbreit1}
\end{figure}
This is done in figure \ref{fig:nbreit1} where the mean charged
multiplicity from $e^+e^-$ reactions are compared with twice the mean charged
multiplicity of $e p$ reactions~\cite{ZHbreit} in the Breit frame. The
results are plotted
as function of the photon virtuality in case of $e p$ and for $e^+e^-$, its
center of mass energy. The good agreement between the two reactions indicates
that the electron side in $e p$ reactions behaves similarly to that in
$e^+e^-$.

\subsection{Multiple interactions}

While one could learn about hadronization from $e^+e^-$ reactions as far as
the electron side of $ep$ is concerned, the situation at the proton side is
different. For very low $Q^2$, the quasi--real photon shows structure
similar to a hadron. The resolved photon shows a clear spectator
structure~\cite{ZEUSrem} just like in the case of the proton where one has
its remnant. One thus can have situations where the remnants of the photon
and proton might also interact, thus producing events with multiple
interactions~\cite{Sjmi}.

\setlength{\unitlength}{0.7mm}
\begin{figure}[hbtp]
\begin{picture}(100,110)(0,1)
\mbox{\epsfxsize8.0cm\epsffile{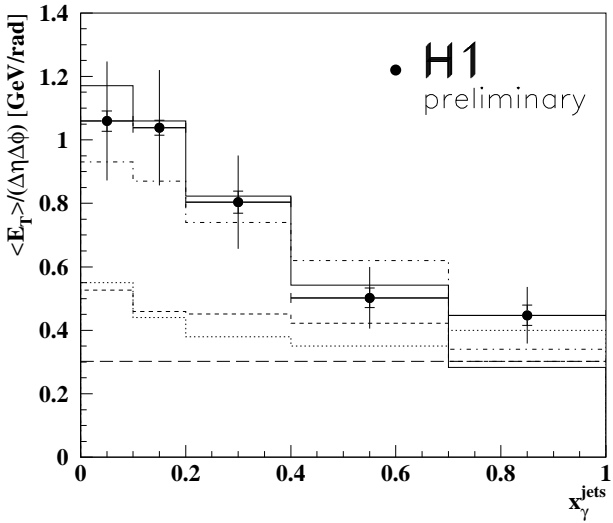}}
\end{picture}
\caption{Transverse energy flow outside the two jets as function of
$x_{\gamma}$. The histograms which are closer to the data points are MC
generators which include multiple interactions, while the lower histograms
don't have interactions of the beam remnants.}
\label{fig:h1mi}
\end{figure}

The H1~\cite{H1mi} collaboration studied photon proton interactions producing
two high $E_T$ jets. They chose the central photon--proton
collision region to look for transverse energy flow outside the two jets
with the highest $E_T$. This flow is compared in figure \ref{fig:h1mi} with
predictions of Monte Carlo generators with and without the multiple
interaction option. This comparison indicated that multiple interactions are
needed in order to be able to reproduce the behaviour of the data.

\setlength{\unitlength}{0.7mm}
\begin{figure}[hbtp]
\begin{picture}(100,130)(0,1)
\mbox{\epsfxsize7.0cm\epsffile{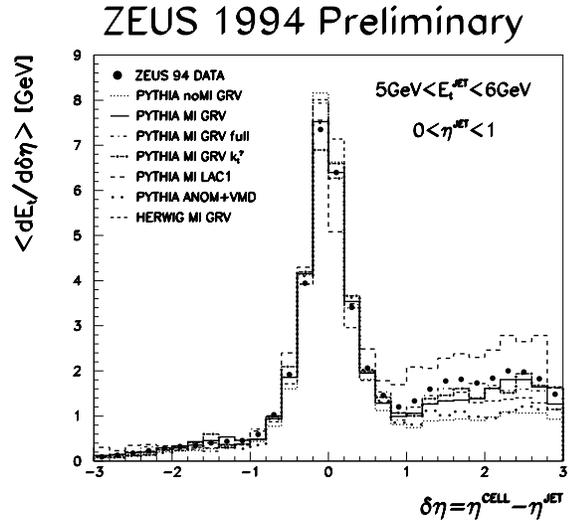}}
\end{picture}
\vspace{-1.4cm}
\caption{ The uncorrected transverse energy flow around the jet axis of jets
in the $E_T$ range of 5--6 GeV (dots) compared to predictions of Monte Carlo
generators (histograms) using some of the available options.}
\label{fig:jetshape}
\end{figure}

However, this conclusion is not unique. The flexibility of the generators
include many more parameters which can be tuned to produce a similar effect
to multiple interactions. This is shown by the ZEUS
collaboration~\cite{ZEUSmi} in figure \ref{fig:jetshape}, where the
uncorrected transverse energy flow around the jet axis is compared with Monte
Carlo predictions using different options and different parametrizations of
the photon structure function. One can choose a combination of these
parameters which does not include multiple interactions and nevertheless get a
fair description of the data.  It is thus clear that in order to have a
better understanding of the fragmentation in the proton region one needs a
more detailed study and further tuning of the different generators.

We can summarize this chapter on fragmentation and hadronization studies
with the statement that the general features of jet fragmentations,
hadronizations, particle multiplicities are quite well described by the
generators. One needs much effort to do systematic tuning for finer details.
One worrysome problem is that the large $p_T$ distribution of final state
hadrons is not properly described. For HERA, the generators need further
tuning for the description of the proton side, with a possible sign for
multiple interactions.

\section{Interplay of soft and hard processes}

\subsection{Operational definition}

It is not completely clear what one means by soft and hard interaction.
One would have hoped that by going to the region of DIS one has a better
way of probing the hard interactions. As a guideline to help distinguish
the two, let us define some operational criteria for what we would
consider as a soft and as a hard process. We can not do it in the most
general terms, but let us concentrate on some selected measurements:
total cross sections and elastic cross sections, the first being the most
inclusive and the latter the most exclusive measurement we can make. At
high energies, both these processes are dominated by a pomeron exchange.

As discussed earlier, the total $\pi^{\pm}p, K^{\pm}p, pp, \bar{p}p$ and
$\gamma p$ cross sections show a slow dependence on the center of mass
energy $W$, consistent with the so--called soft pomeron~\cite{DL}, having a
trajectory
\begin{equation}
\alpha_{\pom(\rm soft)} = 1.08 + 0.25 t
\end{equation}
The hard or the perturbative pomeron, also called the Lipatov pomeron or the
BFKL~\cite{BFKL} pomeron, is expected to have a trajectory
\begin{equation}
\alpha_{\pom(\rm hard)} = 1 + \frac{12 \ln 2}{\pi}\alpha_S
\end{equation}
The definition of the hard $\pom$ is quite vague. First, the value of the
intercept which is usually taken as 1.5 is a very rough estimate using the
expression of the expected power of the reggeized gluon. Using a leading
order calculation in $\ln 1/x$, the distribution of the momentum density of
the gluon is expected to have the form $xg(x,Q^2) \sim x^{-\lambda}$ where
$\lambda=\alpha_S/0.378$. Although usually the value of $\lambda$ is taken to
be 0.5~\cite{K0.5}, one should note that this requires a value of
$\alpha_S=0.18$, which happens only at large $Q^2$, whereas the BFKL
calculation is expected to be valid for moderate $Q^2$ values. The second
comment about the assumed hard $\pom$ form is the fact that the slope of this
trajectory is taken to be zero. The reason for this assumption can be
understood intuitively by the fact that the slope is inversely proportional
to the average transverse momentum square of hadrons, which is expected to be
much larger in hard interactions compared to soft ones.

Following the above definitions of the soft and the hard pomeron, we have
some expectations for the behaviour of the total $\gamma^*p$ cross
section, $\sigma_{tot}^{\gamma^*p}$, and the elastic one, which in the
HERA case is the reaction $\sigma(\gamma^*p\to Vp)$:

\hspace{-1cm}
\begin{tabular}{l|l|l|l} \hline \hline
quantity & $W$ dep & soft & hard \\ \hline
$\sigma_{tot}^{\gamma^*p}$ & $(W^2)^{\alpha_{\pom}(0)-1}$ & $(W^2)^{0.08}$ &
$(W^2)^{0.5}$ \\ \hline
slope $b$ of $\frac{d\sigma}{dt}$ & $\sim 2\alpha'\ln W^2$ & shrink. & no
shrink. \\ \hline
$\sigma(\gamma^*p\to Vp)$ & $\sigma_{tot}^2/b$ & $(W^2)^{0.16}/b $ & $(W^2)^1$
\\ \hline \hline
\end{tabular}

\vspace{3mm}

We thus turn now to the recent HERA data to check their energy behaviour.

\subsection{The total $\gamma^*p$ cross section, $\sigma_{tot}^{\gamma^*p}$}

The total $\gamma^*p$ cross section, $\sigma_{tot}^{\gamma^*p}$, can be
related to the proton structure function $F_2$ through the relation
\begin{equation}
F_2(x,Q^2) = \frac{Q^2(1 - x)}{4\pi^2\alpha}\frac{Q^2}{Q^2 +
4m_p^2x^2}\sigma_{tot}^{\gamma^*p}(x,Q^2)
\label{sigf2}
\end{equation}
where the total $\gamma^*p$ includes both the cross section for the
absorption of transverse and of longitudinal photons. In this expression
the Hand~\cite{hand} definition of the flux of virtual photons is used.

DL extended~\cite{DLext} their parametrization of the total photoproduction
cross section also for virtual photons, using however the same energy
behaviour as for the real photon case. The ALLM~\cite{ALLM} parametrization
assumed that the power of the energy behaviour, $\Delta$ (same meaning as the
$\epsilon$ used by DL), has a $Q^2$ dependence with a smooth transition from
the value of 0.05 at $Q^2$=0 to $\sim$ 0.4 at high $Q^2$. A similar
assumption is made also by CKMT~\cite{CKMT}, who assume a somewhat
simpler $Q^2$ dependence for $\Delta$. Badelek and Kwiecinski~\cite{BK} take
a linear combination of generalized vector dominance model (GVDM) and a QCD
based calculation, where the weight of each part is $Q^2$ dependent.

\setlength{\unitlength}{0.7mm}
\begin{figure}[hbtp]
\begin{picture}(100,145)(0,1)
\mbox{\epsfxsize7.8cm\epsffile{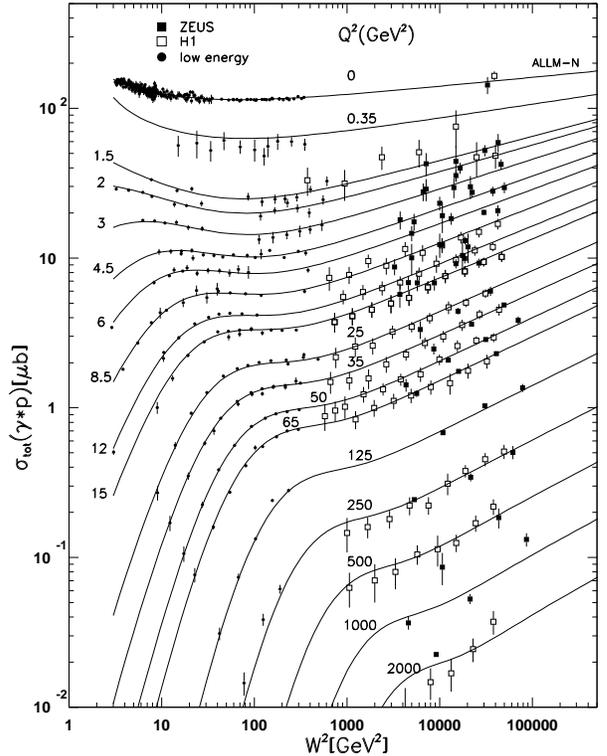}}
\end{picture}
\caption{The total $\gamma^*p$ cross section as function of $W^2$ from the
$F_2$ measurements for different $Q^2$ values. The
lines are the expectations of a new ALLM type parametrization.}
\label{fig:gstarp_allm}
\end{figure}

Figure \ref{fig:gstarp_allm} presents the dependence of
$\sigma_{tot}^{\gamma^*p}$, obtained through equation \ref{sigf2} from the
measured $F_2$ values~\cite{ZEUSf2,H1f2}, on the square of the center of mass
energy, $W^2$, for fixed values of the photon virtuality $Q^2$. While the
data below $Q^2$=1 GeV$^2$ show a very mild $W$ dependence, the trend
changes as $Q^2$ increases. Note that for higher values of $Q^2$ one sees the
typical threshold behaviour for the case when $W^2<Q^2$~\cite{LM}. The curves
are the results of a new ALLM type parametrization which added to the
earlier data used in the previous fit data from E665~\cite{E665}, new data
from NMC~\cite{newNMC} and the published HERA~\cite{pubHERA} data.

\setlength{\unitlength}{0.7mm}
\begin{figure}[hbtp]
\begin{picture}(100,125)(0,1)
\mbox{\epsfxsize6.5cm\epsffile{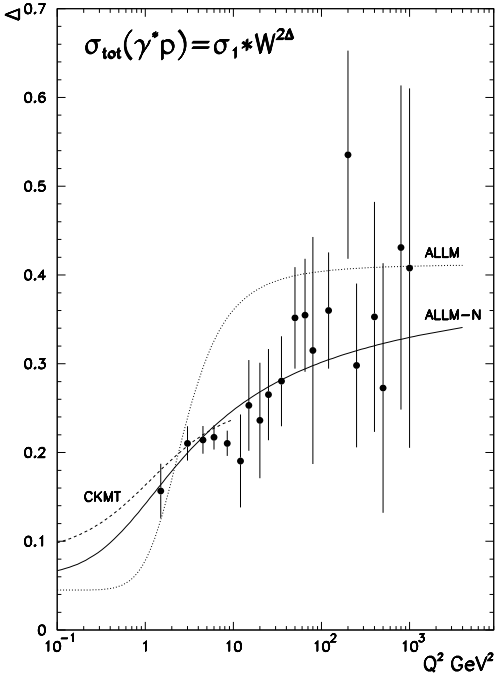}}
\end{picture}
\caption{The $Q^2$ dependence of the parameter $\Delta$ obtained from a fit
of the expression $\sigma_{tot}^{\gamma^*p} = \sigma_1 W^{2\Delta}$ to the
data in each $Q^2$ bin. The curves are the expectations of the
parametrizations mentioned in the text.}
\label{fig:delta}
\end{figure}

In order to see how the slope of the $W$ dependence changes with $Q^2$,
the cross section values in the region where $W^2\gg Q^2$ were
fitted~\cite{AL} to the form $\sigma_{tot}^{\gamma^*p} = \sigma_1
W^{2\Delta}$ for each fixed $Q^2$ interval. The resulting values of
$\Delta$ from the fit are plotted against $Q^2$ in figure
\ref{fig:delta}. The curves are the expectations of the CKMT
(parametrization valid~\cite{CKMT} only up to $Q^2 \sim$ 5--10 GeV$^2$),
the ALLM (fitted without the HERA data) and the updated ALLM--N
parametrization, which includes also some of the recent HERA data in its
fit.  One can see the slow increase of $\Delta$ with $Q^2$ from the value
of 0.08 at $Q^2$=0, to around 0.2 for $Q^2 \sim$ 10--20 GeV$^2$ possibly
followed by a further increase to around 0.3--0.4 at high $Q^2$. One
would clearly profit from more precise data, expected to come soon.

We can thus conclude that the energy behaviour of $\sigma_{tot}^{\gamma^*p}$
indicates that there is a smooth transition from soft to hard interactions
with an interplay between the two in the intermediate $Q^2$ range.

\subsection{Vector meson production in $\gamma p$ and in $\gamma^* p$}

Given the behaviour of the $\sigma_{tot}^{\gamma^*p}$ data, what kind of
energy behaviour would one expect for the `elastic' process $\gamma^*p\to
Vp$ for real and virtual photons? In case of photoproduction we have seen
that the total cross section follows the expectations of a soft DL type
$\pom$. Thus if one takes into account the shrinkage at the HERA energies,
one expects $\sigma(\gamma p\to Vp) \sim W^{0.22}$. In case of DIS
production of vector mesons in the range $Q^2 \sim$ 10--20 GeV$^2$, the
expectations are  $\sigma(\gamma^*p\to Vp) \sim W^{0.8}$, since in this case
one expects almost no shrinkage.

\setlength{\unitlength}{0.7mm}
\begin{figure}[hbtp]
\begin{picture}(100,120)(0,1)
\mbox{\epsfxsize7.5cm\epsffile{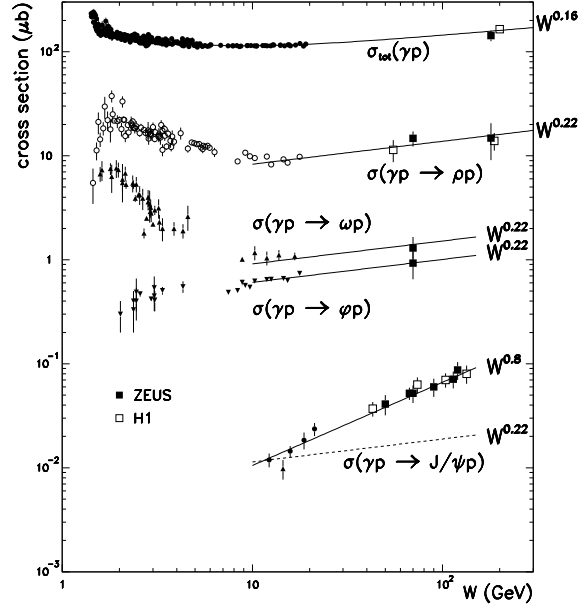}}
\end{picture}
\caption{The total and `elastic' vector meson photoproduction measurements
as function of $W$, for the vector mesons $\rho, \omega, \phi$ and $J/\Psi$.
The curve to the total photoproduction cross section is the DL
parametrization ($W^{0.16}$). The other lines are curves of the form
$W^{0.22}$ and $W^{0.8}$.}
\label{fig:phpvm}
\end{figure}

Figure \ref{fig:phpvm} presents the measurements of the total and
`elastic' vector meson photoproduction cross sections as function of the
$\gamma p$ center of mass energy $W$. As one can see, the high energy
measurements of the total and the $\rho$~\cite{ZEUSrho,H1rho}, $\omega$
and $\phi$ photoproduction~\cite{ZEUSrho} follow the expectations of a
soft DL type pomeron. However, the cross section for the reaction $\gamma
p \to J/\Psi p$~\cite{ZEUSpsi,H1psi} rises much faster than the expected
$W^{0.22}$ rise from a soft reaction. In fact, it can be well described by
a power behaviour of $\sim W^{0.8}$. This surprising behaviour can be
understood if one considers the scale which is involved in the
interaction. In case of photoproduction reaction, the scale cannot be set
by the photon since $Q^2=0$. The scale is set by the mass of the vector
meson and by the transverse momentum involved in the reaction. Thus, for
the lighter vector mesons the scale is still small enough to follow a soft
behaviour. However, the mass of the $J/\Psi$ is large enough to produce a
scale which would be considered as a hard interaction.

The reaction $\gamma^* p \to \rho^0 p$ has been measured~\cite{ZEUSdrho,H1drho}
at two $Q^2$ values of 8.8 and 16.9 GeV$^2$. The reaction  $\gamma^* p
\to \phi p$ has been measured~\cite{ZEUSdrho} for $Q^2$ of 8.3 and 14.6
GeV$^2$. Since in these cases the scale is set by $Q^2$, one expects the
energy behaviour to be $\sim W^{0.8}$.
\setlength{\unitlength}{0.7mm}
\begin{figure}[hbtp]
\begin{picture}(100,110)(0,1)
\mbox{\epsfxsize8.0cm\epsffile{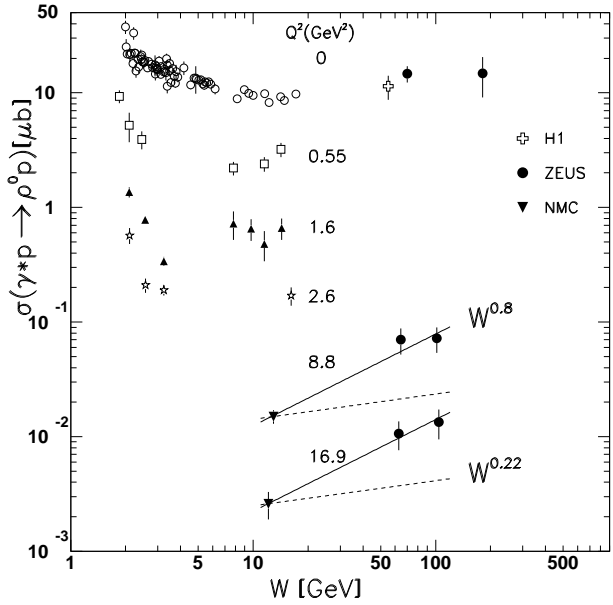}}
\end{picture}
\caption{The dependence of the cross section for the reaction $\gamma^* p
\to \rho^0 p$ on $W$, for different $Q^2$ values.}
\label{fig:disrho}
\end{figure}
The measurements from HERA are compared to the lower energy ones from
NMC~\cite{NMC} and are displayed in figure \ref {fig:disrho} for the
$\rho^0$ and in figure \ref {fig:disphi} for the $\phi$. In both cases one
can see a faster rise with $W$ than the soft $W^{0.22}$ behaviour and in
good agreement with $W^{0.8}$. These results are thus consistent with the
$Q^2$ dependence of $\Delta$ as shown in figure \ref{fig:delta}.
\setlength{\unitlength}{0.7mm}
\begin{figure}[hbtp]
\begin{picture}(100,110)(0,1)
\mbox{\epsfxsize8.0cm\epsffile{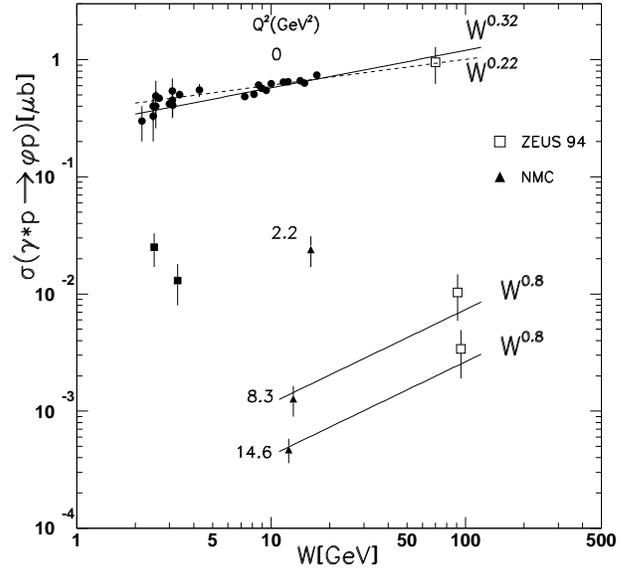}}
\end{picture}
\caption{The dependence of the cross section for the reaction $\gamma^* p
\to \phi p$ on $W$, for different $Q^2$ values.}
\label{fig:disphi}
\end{figure}

The ratio of the $\phi$ to $\rho^0$ cross sections is expected to be 2:9
according to SU(4).
\setlength{\unitlength}{0.7mm}
\begin{figure}[hbtp] \begin{picture}(100,110)(0,1)
\mbox{\epsfxsize8.0cm\epsffile{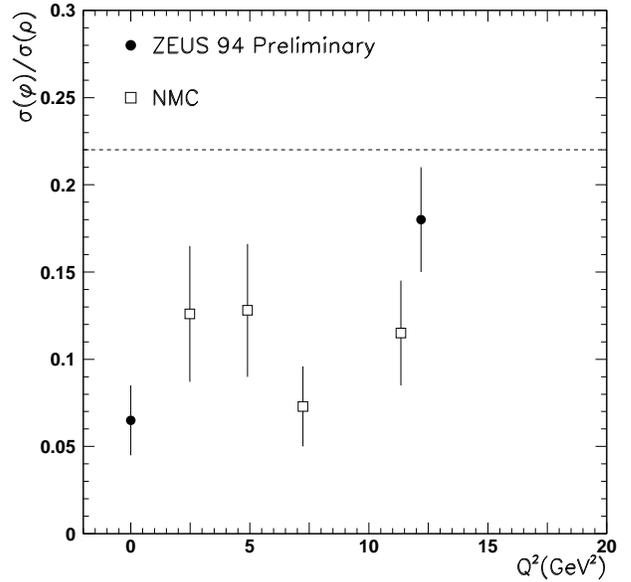}}
\end{picture}
\caption{The ratio of the $\phi$ to $\rho^0$ cross sections as a function of
$Q^2$ for the ZEUS and NMC data. The dotted line is the expected ratio from
SU(4).}
\label{fig:phi2rho}
\end{figure}
However experimentally this ratio is about 0.07 at
$Q^2$=0~\cite{bauer} and increases to about 0.1 at $Q^2 \sim$ 10 GeV$^2$ in
the $W$ range of $\sim$ 10 GeV of the NMC~\cite{NMC} experiment. The
disagreement was explained as a suppression due to the difference in the
masses and wavefunctions of the two vector mesons. However, at higher $W$
and higher $Q^2$, the effect of these differences are expected~\cite{halfms}
to be negligible and the ratio should reach the value of 2/9. One of the
interesting questions is whether it is enough to go to higher $W$ or does
one need both $W$ and $Q^2$ to be large.

The ZEUS~\cite{ZEUSdrho} collaboration has measured this ratio both at
$Q^2$=0 and at $Q^2 \sim$ 15 GeV$^2$, at a value of $W \sim$ 100 GeV.  These
measurements, together with those of NMC, are plotted in figure
\ref{fig:phi2rho}. It is clear from the figure that in order to reach the
expected value of 2/9 one needs both $W$ and $Q^2$ to be large. One also sees
that the value is still not reached, thereby hinting that we still are in the
region where the interplay between soft and hard interactions is observed.

Before continuing to the next chapter it is worthwhile to note that the
properties observed for vector mesons have a natural explanation in
QCD, where vector meson production with a large scale can be described by
an exchange mechanism of a pomeron consisting of two gluon. For example, in
the case of the model of Brodsky et al.~\cite{brodsky} one expects that the
differential $\rho^0$ cross section produced by longitudinal photons should
be proportional to the gluon distribution in the proton:
\begin{equation}
\frac{d\sigma}{dt}(\gamma_L^* p \to \rho^0 p) \sim \frac{[\alpha_S(Q^2) x
g(x,Q^2)]^2}{Q^6}C_{\rho}
\end{equation}
Since at low $x$ values $[\alpha_S(Q^2) x g(x,Q^2)]^2 \sim Q$ and since
the $k_T$ dependence of the $\rho^0$ wave function introduces~\cite{fks}
another
$Q^{0.5}$ dependence, the expectations of the QCD calculation are that
the data should have a $Q^n$ dependence, where $n$=4.5--5. The
ZEUS~\cite{ZEUSdrho} experiment finds $n$=4.2$\pm$0.8$^{+1.4}_{-0.5}$ and
the H1~\cite{H1drho} experimental result is $n$=4.8$\pm$0.8 (statistical
error only). The $x$ dependence of the ZEUS~\cite{ZEUSxrho} measurement
is consistent with their gluon determination from their $F_2$
measurement.

The conclusion of this chapter is that the reactions $\gamma p \to V p$ and
$\gamma^* p \to V p$ are consistent with the $W$ behaviour of
$\sigma_{tot}^{\gamma^*p}$. According to the above operational definition,
we observe a transition from soft to hard interactions, with the results in
the presently measured kinematic range being driven by an interplay of both.

When a large scale is present, being the virtuality of the photon or the mass
of the vector meson, the cross section is consistent with a rise driven by
the rise of the gluon momentum density $x g(x,Q^2)$ with $W$. The pomeron
exchange mechanism described by two gluons gives results consistent with the
data. In the next chapter we will see what we can say about the structure of
the pomeron.

\section{ Large rapidity gap events in DIS}

One of the surprising results of HERA was the discovery~\cite{LRGZ,LRGH} of
large rapidity gap (LRG) events in DIS processes. These events were shown
to be consistent with a picture in which the virtual photon $\gamma^*$
diffracts into a system with invariant mass $M_X$, exchanging a
colourless object with the properties expected from a pomeron.

In order to understand why this was surprising, consider photoproduction
processes. As we saw in chapter 2, about 10\% of the total $\gamma p$
cross section comes from processes where the photon diffracts. This
hadronic behaviour of the photon is accepted because of the fact that a
real photon, which is energetic enough, can fluctuate into a hadronic
state before interacting and the fluctuation time is large enough
compared to the interaction time so that the hadronic nature of the
photon is felt in the interaction. The fluctuation time can be expressed
as~\cite{Ioffe} $t_f \sim 2E_{\gamma}/m_V^2$, where $E_{\gamma}$ is the
photon energy and $m_V$ is the mass of the hadronic state into which the
photon fluctuates. The fluctuation time of a virtual photon with
virtuality $Q^2$ is $t_f \sim 2E_{\gamma}/(m_V^2 + Q^2)$ and thus as the
photon virtuality increases, the fluctuation time decreases and the
virtual photon behaves like a point--like object. Therefore the
generators describing DIS processes did not include diffraction as one of
the possible reactions.

However, a closer look at the above argument shows that in fact the
diffraction of virtual photons was to be expected. In the region of high
W, when $W^2 \gg Q^2$, the fluctuation time can be
expressed~\cite{Ioffe1} as $t_f \sim 1/(m_p x)$. Thus at low $x$, the
fluctuation time is large enough for a virtual photon to diffract like
any hadron.
\setlength{\unitlength}{0.7mm}
\begin{figure}[hbtp]
\begin{picture}(100,125)(0,1)
\mbox{\epsfxsize9.0cm\epsffile{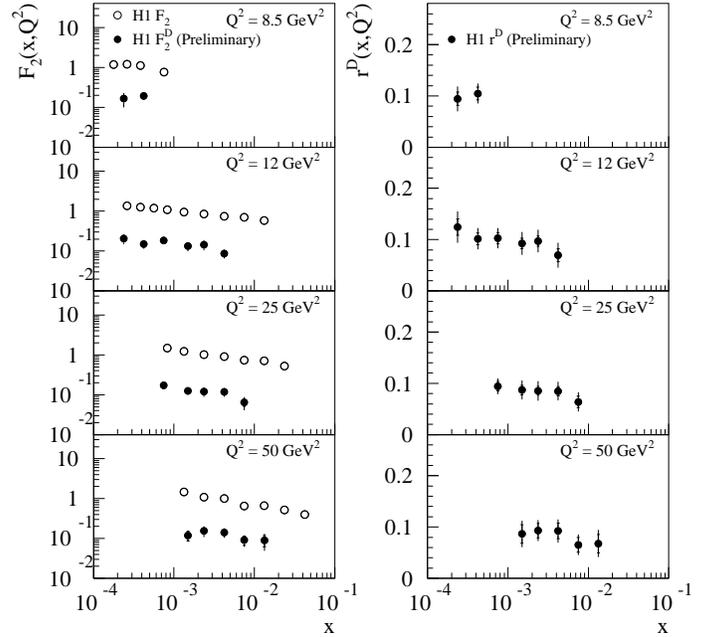}}
\end{picture}
\caption{ The inclusive structure function (open circles) and the
contribution coming from LRG events (full dots) as function of Bjorken
$x$ for different $Q^2$ intervals. In the right part of the figure, the
ratio $r^D$ of the two structure function is plotted as function of $x$
for the same $Q^2$ intervals as in the left part.}
\label{fig:lrgratio}
\end{figure}
The rate of these LRG events is about 10\% of the total cross
section~\cite{LRGZ,LRGH} and is displayed in figure \ref{fig:lrgratio}
for data of the H1~\cite{LRGH1} experiment. The ratio of the LRG events
seems to be in first order independent on $x$ and on $Q^2$ and thus are a
leading twist effect.

In order to define a structure function of the LRG events, it is useful
to define the following variables in addition to $x$ and $Q^2$: the four
momentum transfer squared $t$ at the proton vertex, the fraction
$x_{\pom}$ of the proton momentum carried by the (generic) pomeron emitted
by the proton and the fraction $\beta$ of the pomeron momentum carried by
the struck quark. The last two variables can be expressed as:
\begin{equation}
x_{\pom} = \frac{M_X^2 + Q^2}{W^2 + Q^2}, \ \ \ \ \ \beta =
\frac{Q^2}{M_X^2 + Q^2}
\end{equation}
The three fractional momenta are related by $x = x_{\pom} \beta$. Since
$t$ is not yet measured by the experiments, one integrates and defines a
threefold differential cross section in the form
\begin{equation}
\frac{d^3\sigma}{d\beta dQ^2 dx_{\pom}} = \frac{2\pi \alpha^2}{\beta
Q^2}[1 + (1-y)^2] F_2^{D(3)}(\beta, Q^2, x_{\pom})
\end{equation}
In this expression we have neglected the contribution coming from the
longitudinal part of the structure function.

\subsection{Factorization}

Assuming that the LRG events are produced in a diffractive process where a
pomeron is exchanged, it would be interesting to check what are the
properties of this exchanged object. Is it an Ingelman--Schlein(IS)~\cite{IS}
type of pomeron which behaves similar to a regular hadron? Is it a pomeron as
modelled by Nikolaev--Zakharov~\cite{NZ} by two or more gluon exchanges?  In
the former case, the $F_2^{D(3)}$ structure function can be factorized into a
part which is the flux $f_{\pom}(x_{\pom})$ of pomerons emitted by the
proton, multiplied by the structure function of the pomeron,
$F_2^{\pom}(\beta, Q^2)$. In the latter case, there is an explicit
factorization breaking in the model. Therefore, in order to check the
factorization assumption, the measured structure function $F_2^{D(3)}$ was
fitted in different $\beta, Q^2$ bins to the form
\begin{equation}
F_2^{D(3)} = \left( \frac{1}{x_{\pom}} \right)^a C_{\beta,Q^2}
\end{equation}
where $a$ is  kept fixed for all $\beta, Q^2$ bins and $C$ is a constant
allowed to vary from bin to bin. The form chosen for the flux factor is
driven by the Regge phenomenology where one expects the flux of pomerons
to have the form
\begin{equation}
f_{\pom}(x_{\pom}) = \left( \frac{1}{x_{\pom}} \right)^{2\alpha_{\pom}(t)
- 1}
\label{fact}
\end{equation}
\setlength{\unitlength}{0.7mm}
\begin{figure}[hbtp]
\begin{picture}(100,130)(0,1)
\mbox{\epsfxsize8.0cm\epsffile{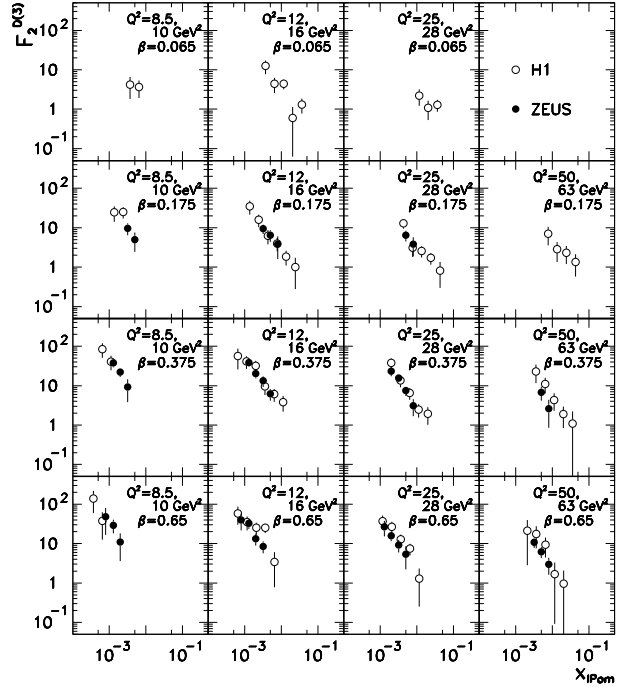}}
\end{picture}
\caption{$F_2^{D(3)}(\beta, Q^2, x_{\pom})$ as a function of $x_{\pom}$
for fixed $\beta, Q^2$ bins. }
\label{fig:f2d3_zh1}
\end{figure}

The measured $F_2^{D(3)}$ by H1~\cite{H1fact} and ZEUS~\cite{Zfact} are
plotted in figure \ref {fig:f2d3_zh1} as function of $x_{\pom}$ for
fixed $\beta, Q^2$ bins and fitted to the form expressed in equation \ref
{fact}. The data are consistent with the factorization hypothesis,
yielding for the power $a$ of the flux factor the result
\begin{eqnarray}
 a_{H1} & = & 1.19~\pm~0.06~(stat)~\pm~0.07~(syst) \nonumber \\
 a_{ZEUS} & = & 1.30~\pm~0.08~(stat)~^{+~0.08}_{-~0.14}~(syst)
\label{power}
\end{eqnarray}
The expected power for a DL type soft pomeron would be about 1.10--1.12,
depending on the assumed slope of the diffractive peak used in the
integration over $t$ and on the value of the slope of the pomeron trajectory.
For the case of a hard pomeron as described in the earlier chapter one
expects $a$ to be 2. The result is consistent with that of a soft pomeron but
also with that of an admixture of both soft and hard pomeron as discussed
above. This could be another example of the interplay of soft and hard
interactions.

Two notes in passing: (1) though the result is consistent with
factorization, it can also be described by the Nikolaev--Zakharov model
which assumes factorization breaking. However the breaking in the measured
kinematic region is too small to be distinguished by the present accuracy
of the data; (2) the behaviour of the data can also be explained by a
model like that of Buchm\"uller and Hebecker~\cite{buch} where the
concept of the pomeron is not used and instead assumes a
non--perturbative colour cancellation induced by wee partons.

\subsection{The structure of the pomeron}

If the pomeron has substructure, as postulated by the Ingelman--Schlein
model, one could learn about the parton of the pomeron by using the
$F_2^{D(3)}$ structure function and compare it to different shapes of
parton distribution functions. From the shapes of the distribution of
some kinematical variables the HERA experiments found that one needs
quarks which have a hard distribution of the form $\beta (1-\beta)$. If
in addition one assumes that the partons in the pomeron satisfy a
momentum sum rule like a regular hadron, one can get the coefficient of
the parton distribution.  This is shown in figure \ref{fig:f2dnorm},
where the pomeron is assumed to contain quarks having a hard momentum
distribution and the momentum sum rule is assumed to hold. The comparison
to the data is done assuming the DL or the IS form for the flux of
pomerons. One can see that independent of the assumed flux, the momentum
of the pomeron is not saturated by quarks alone.
\setlength{\unitlength}{0.7mm}
\begin{figure}[hbtp]
\begin{picture}(100,100)(0,1)
\mbox{\epsfxsize8.0cm\epsffile{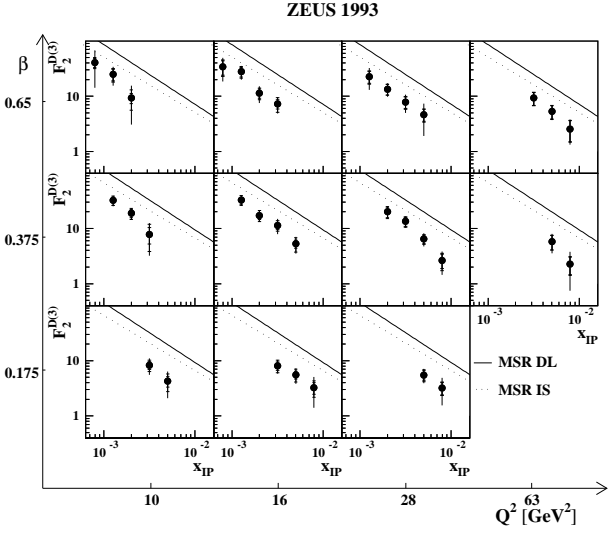}}
\end{picture}
\caption{Comparison of the structure function $F_2^{D(3)}$ with the
momentum sum rule, using the DL(full line) or the IS(dotted line) flux
factors.}
\label{fig:f2dnorm}
\end{figure}

Another way of studying the partonic structure of the pomeron is to
perform an experiment similar to the two jet production associated with a
tagged leading proton experiment carried out by UA8~\cite{UA8}.
\setlength{\unitlength}{0.7mm}
\begin{figure}[hbtp]
\begin{picture}(100,120)(0,1)
\mbox{\epsfxsize8.0cm\epsffile{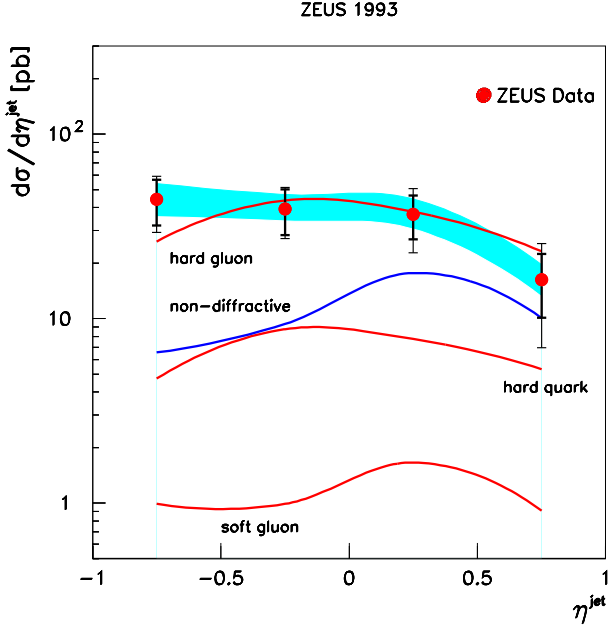}}
\end{picture}
\caption{The inclusive jet photoproduction cross section, compared with
different parton distributions and with expectations from non--diffractive
background, as described in the text.}
\label{fig:phpdif}
\end{figure}
This was
done by ZEUS~\cite{Z2jd} which studied large $p_T$ jets photoproduced
with a LRG. The cross section values of these events are compared in figure
\ref
{fig:phpdif} to different parton distributions in the pomeron, as well as
with background coming from non--diffractive events.
The flux of the pomeron was assumed to be given by DL and the pomeron was
assumed to consist of either only quarks or only gluons, soft or hard, in
each case saturating the momentum sum rule. As can be seen from the
figure, the data can be explained either by a pure hard gluonic component
or by a combination of hard quarks and gluons in the pomeron.

\setlength{\unitlength}{0.7mm}
\begin{figure}[hbtp]
\begin{picture}(100,120)(0,1)
\mbox{\epsfxsize8.0cm\epsffile{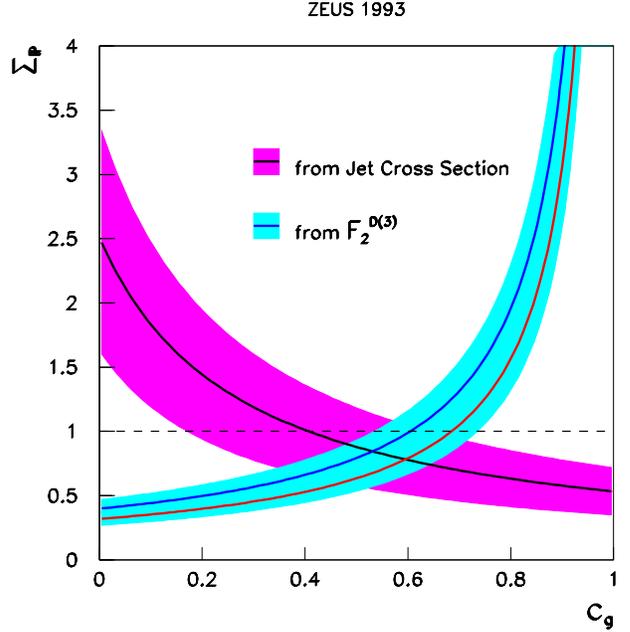}}
\end{picture}
\caption{ The normalization $\Sigma_{\pom}$ needed to describe the DIS
and the photoproduction data as a function of the fraction $c_g$ of gluons
in the pomeron.}
\label{fig:sumrule}
\end{figure}
The conclusions drawn from the comparisons in figure \ref {fig:f2dnorm} and
from those in figure \ref {fig:phpdif}, when considered separately, depend on
the choice of the pomeron flux and on the assumption of the validity of a sum
rule for the pomeron. One can however try and do a comparison based on the
joint results of the DIS and the photoproduction data, leaving as free
parameters the amount of gluons, $c_g$, and that of quarks, $1-c_g$, in the
pomeron as well as the overall normalization $\Sigma_{\pom}$ needed to
reproduce the results of the two measurements.  The region where
$\Sigma_{\pom}$ is the same for both determines the amount of gluons in the
pomeron in a model independent way, provided the same pomeron is involved in
both processes. Such a comparison is shown in figure \ref {fig:sumrule}, from
which one can conclude that the gluon carries a fraction of $0.3<c_g<0.8$ of
the pomeron momentum. This result is independent of the assumed expression
for the flux, of the validity and of the assumption of a momentum sum rule
for the pomeron.

\subsection{LRG between jets}

Another way of looking for processes where the pomeron is expected to
behave differently than the soft pomeron is in large $t$ processes. One
possibility is to look for gaps between high $p_T$ jets, which corresponds
to a high $t$ two-body reaction. Study of the rate of jets as function of
the rapidity gap between the jets can be a source of observing the
presence of pomeron exchange in hard processes.

\setlength{\unitlength}{0.7mm}
\begin{figure}[hbtp]
\begin{picture}(100,110)(0,1)
\mbox{\epsfxsize8.0cm\epsffile{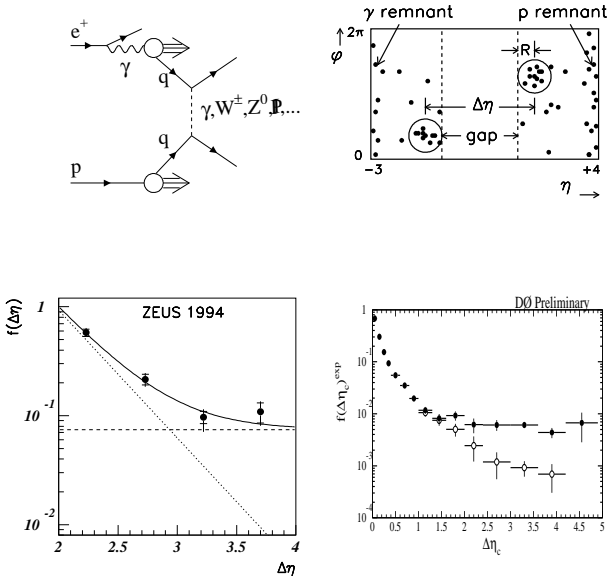}}
\end{picture}
\caption{ (a) Resolved photoproduction via a colour singlet exchange. (b)
The rapidity gap morphology of a two jets event, as described in the
text. (c) The gap fraction $f$ as function of the pseudorapidity interval
$\Delta \eta$ for the ZEUS data, compared with the result of a fit to an
exponential plus a constant. (d) The gap fraction between two jets for
the D0 data.}
\label{fig:jetgap}
\end{figure}
Figure \ref {fig:jetgap}(a) describes schematically an example of colour
singlet exchange in resolved photoproduction in which a parton in the photon
scatters from a parton in the proton, via $t$--channel exchange of a colour
singlet object. The event morphology is described in figure \ref
{fig:jetgap}(b) where two jets in the final state are shown as circles in
($\eta,\phi$) space and $\Delta \eta$ is defined as the distance in $\eta$
between the centers of the two jet cones. The gap fraction $f(\Delta \eta)$
for the ZEUS~\cite{Zcs} data are compared in figure \ref {fig:jetgap}(c) with
the result of a fit to an exponential plus a constant from which one
concludes that there is an excess of about 0.07 in the gap fraction over the
expectation from colour non--singlet exchange.

There were indeed indications from the CDF~\cite{CDFlrg} and the
D0~\cite{D0} collaboration for the existence of rapidity gap between jets
at a rate of about 0.01, as shown for the D0 data in figure \ref
{fig:jetgap}(d). The ZEUS collaboration measurement of 0.07 can be
compatible with the Tevatron result if one takes into account the fact
that the survival probability of a gap is larger in $\gamma p$ reactions
than in $\bar{p} p$ ones.

To conclude this chapter we can say that large rapidity gaps have been
observed also in DIS processes. They can be interpreted as being due to
diffraction, with the exchange of a pomeron. The diffractive structure
function is consistent with the factorization hypothesis in the kinematic
range where data exist. In a picture of an Ingelman--Schlein type
pomeron with a partonic substructure, a large fraction of the pomeron
momentum is carried by gluons. More detailed study is needed before
concluding about the soft or hard nature of this pomeron.

There are first signs for the existence of a colour singlet exchange at
large $t$.

\section{Summary}

One would like to separate soft from hard interactions. However nothing is
as soft as we would like nor as hard as we would like. There is an
interplay of soft and hard processes at all values of $Q^2$. As $Q^2$ or any
other scale increases, the amount of hard processes seems to increase. In
order to resolve the hard processes one needs a good understanding of the
soft fragmentation and hadronization. By combining various reactions one
can try and extract the perturbative QCD part and to learn more about the
interplay.

\section*{Acknowledgments}

I would like to thank H.Abramowicz for her help in preparing this talk and
this manuscript. The discussions with L.Frankfurt, E.Gotsman, U.Maor and
A.K.Wroblewski were very useful. Thanks also to R.Klanner for a careful
reading of this manuscript.

\end{document}